# Mobile Edge Cloud: Opportunities and Challenges

Sayed Chhattan Shah
Mobile Grid and Cloud Computing Lab
Department of Information Communication Engineering
Hankuk University of Foreign Studies, South Korea
shah@hufs.ac.kr
www.mgclab.com

*Abstract*— Mobile edge cloud is emerging as a promising technology to internet of things and cyber physical system applications such as smart home and intelligent video surveillance. In smart home, various sensors are deployed to monitor the home environment and physiological health of individuals. The data collected by sensors are sent to an application, where numerous algorithms for emotion and sentiment detection, activity recognition and situation management are applied to provide healthcare- and emergency-related services and to manage resources at the home. The executions of these algorithms require a vast amount of computing and storage resources.

To address the issue, the conventional approach is to send the collected data to an application on an internet cloud. This approach has several problems such as high communication latency, communication energy consumption and unnecessary data traffic to the core network. To overcome the drawbacks of the conventional cloud-based approach, a new system called mobile edge cloud is proposed. In mobile edge cloud, multiple mobile and stationary devices interconnected through wireless local area networks are combined to create a small cloud infrastructure at a local physical area such as home.

Compared to traditional mobile distributed computing systems, mobile edge cloud introduces several complex challenges due to the heterogeneous computing environment, heterogeneous and dynamic network environment, node mobility, and limited battery power. The real time requirements associated with internet of things and cyber physical system applications make the problem even more challenging. In this paper, we describe the applications and challenges associated with design and development of mobile edge cloud system and propose an architecture based on a cross layer design approach for effective decision making.

*Keywords - Mobile cloud, Mobile ad hoc cloud, Local cloud, Edge computing, Fog computing.*

## I. INTRODUCTION

Cloud computing provides ubiquitous, convenient and on-demand network access to a shared pool of configurable computing resources [1]. Cloud computing systems include powerful computing resources connected through high speed networks [2]. Due to latest developments in mobile computing and networking technologies it has become feasible to integrate various mobile devices such as robots, aerial vehicles, sensors, and smart phones with cloud computing systems. The approaches for integrating mobile devices with cloud computing systems are divided into two main categories: mobile ad hoc cloud [2] and mobile infrastructure-based cloud [2-7].

In mobile ad hoc cloud, multiple mobile devices interconnected through a mobile ad hoc network are combined to create a virtual supercomputing node. Whereas in mobile infrastructure-based cloud, mobile devices are integrated with a cloud computing system through an infrastructure-based communication network such as cellular network. The integration of mobile devices with cloud computing systems enable mobile devices to access vast amount of processing power and storage space. This makes possible to execute data and computationally intensive applications such as image and video processing on mobile devices. Data storage and execution of such applications on cloud also improve reliability and extend the battery life of mobile devices [2]. Mobile ad hoc cloud computing systems are designed for mobile ad hoc environments whereas mobile infrastructure-based cloud computing systems face issues such as high communication latency, transmission energy consumption, and unnecessary data traffic to the core network.

To overcome the drawbacks of traditional mobile distributed computing systems, a mobile edge cloud is proposed. In mobile edge cloud, multiple mobile and stationary devices interconnected through ad hoc and infrastructure-based local area networks are combined to create a small cloud infrastructure at a local physical area such as home. Mobile edge cloud system aims to serve as a distributed computing and networking infrastructure for several internet of things and cyber physical system applications [2] that require a vast amount of computing and storage resources.

Compared to traditional mobile distributed computing systems [3-7], mobile edge cloud introduces several complex challenges due to the heterogeneous computing environment, heterogeneous and dynamic network environment, node mobility, and limited battery power. The real time requirements associated with IoT and cyber physical system applications make the problem even more challenging. In this paper, we describe the applications and challenges associated with design and development of mobile edge cloud system and propose an architecture based on a cross layer design approach for effective decision making.

The rest of the paper is organized as follow. Section II describes the applications of mobile edge cloud. Section III discusses the challenges associated with design and development of mobile edge cloud systems. Architecture of mobile edge cloud system is described in Section IV. Section V discusses the related work and conclusion is presented in Section VI.

## II. APPLICATIONS OF MOBILE EDGE CLOUD

The mobile edge cloud would enable applications in numerous areas such as security, disaster relief operation, mobile robotics, and smart environments. For motivation, two key applications are described in this section.

### A. Smart Home Environment

"Several environmental, audio, video, and bio sensors, capable of acquiring vital signs such as blood pressure, temperature, and electro cardiogram, are deployed to observe home environment and physiological health of an individual. The data collected by sensors is sent to an application where numerous algorithms for emotion and sentiment detection, activity recognition and situation management are applied to provide healthcare- and emergency-related services and to manage resources at the home. The execution of these computationally intensive real-time tasks require computing capabilities that go beyond those of an individual sensing and processing devices. In order to realize the objective the sensor nodes, wearable computing devices, smart phones and computing devices available at home can be combined to create a mobile edge cloud" [2].

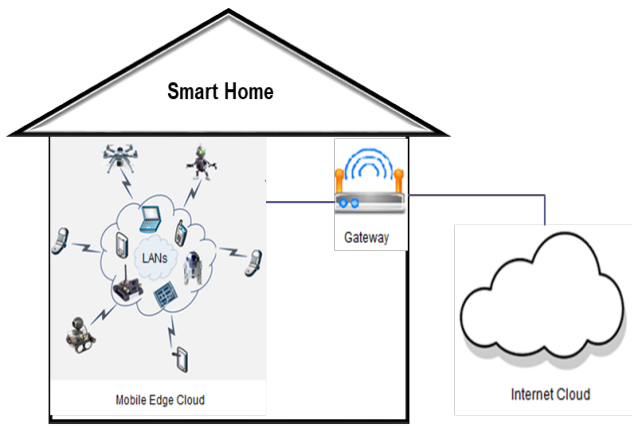

Figure 1: Mobile edge cloud for a smart home

### B. Mobile Intelligent Video Surveillance System

"In order to understand the environment, various mobile robots and micro drones equipped with audio, video and environmental sensors are deployed to collect data which is then processed to construct a three dimensional map of environment in real-time. The collected data is also used to detect and track mobile and stationary targets which involve sophisticated image and video processing algorithms. In order to perform these tasks a vast amount of computing and storage resources are required" [2]. To address the issue, multiple mobile devices such as mobile robots, ground vehicles, unmanned micro aerial vehicles, and wearable devices interconnected through local area networks can be combined to create a real-time and energy efficient mobile edge cloud.

## III. CHALLENGES

Compared to traditional mobile distributed computing systems [3-7], mobile edge cloud introduces several complex challenges due to the heterogeneous computing environment, heterogeneous and dynamic network environment, node mobility, and limited battery power. The real time requirements associated with internet of things and cyber physical system applications make the problem even more challenging. In this section we describe key research challenges. A detailed discussion of challenges common with mobile cloud computing systems is given in [2] and therefore not repeated here.

### A. Heterogeneous Computing Environment

Mobile edge cloud includes several types of mobile and stationary nodes such as sensors, actuators, home appliances, smart phones, and personal computers. Each type of node offers different characteristics and capabilities [4]. This provides opportunities as well as challenges related to resource monitoring, resource scheduling, task distribution and execution, task migration, interoperability, and communication.

### B. Heterogeneous and Dynamic Network Environment

Nodes in mobile edge cloud use different types of wireless ad hoc and infrastructure-based local and wide area networking technologies such as 4G, Wi-Fi, ZigBee, and Bluetooth. It is also common for a node to have multiple communication interfaces, enabling simultaneous communication with one or several devices. Such an environment provides several advantages due to multiple topologies and connections with different characteristics and capabilities. At the same time it also presents numerous challenges related to communication between devices, network and data interoperability, channel types, network interfaces, and network and resource management [3].

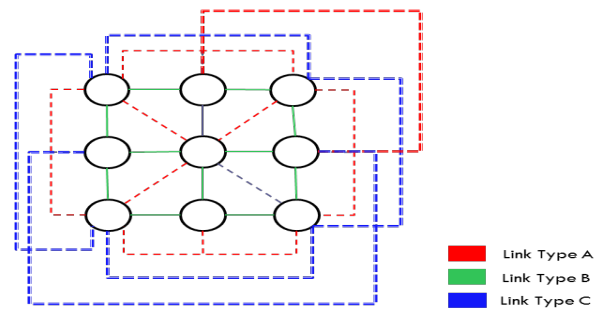

Figure 2: An illustration of heterogeneous network

*C. Node Mobility*

Mobile edge cloud includes both stationary and mobile nodes. Mobile nodes introduce global and local node mobility problems. Global node mobility [13] results in failure of one or several tasks whereas local node mobility [12] [13] may increase data transfer cost or may alter network structure. Task migration or reallocation strategy can be used to manage task failure but this will introduce delay and increase task completion time. Global node mobility also introduces the problem of obsolete information. To address the problem, continuous monitoring of resources is required [10] [12] but it will increase communication cost. An alternate approach is to enquire the status of resources during decision making [16] but this will introduce a delay.

*D. Power Management*

"Key factors that contribute to energy consumption in transmission are the transmission power required to transmit data and the communication cost [11]. Poor design of mobile edge cloud system can significantly increase the communication and energy consumption cost, which limits the life time of nodes and may result in power failure. In the literature, various approaches have been proposed to address the energy consumption problem but they are focused on the conservation of processing energy [8] [10] [14] [15], while saving energy in data transfers between tasks remains an open problem which becomes even more critical for data intensive parallel applications."

*E. QoS Support*

QoS support ensures that application's demands such as energy efficiency, bandwidth guarantees and real-time services are fulfilled. Ensuring the required QoS in presence of node mobility and dynamic and heterogeneous network environment is a challenging task which requires innovative resource management strategies and adaptive routing protocols. Such protocols, for example, should reduce energy consumption either using multi hop communication or transmission power control mechanism if that does not cause tasks to miss the deadline, or send data through multiple routes when network becomes unstable. In addition, the resource management system should actively communicate with routing layer to make effective decisions. In the existing approaches, there is no interaction between routing layer and resource management system [2].

## IV. MOBILE EDGE CLOUD ARCHITECTURE

In mobile edge cloud, multiple mobile and stationary devices interconnected through wireless local area networks are combined to create a small cloud infrastructure at a local physical area. An illustration of mobile edge cloud is given in Figure 3. Mobile edge cloud comprising of multiple mobile and stationary devices interconnected through wireless local area networks is connected to an internet cloud computing system through a gateway node. The gateway node acts as a bridge between mobile edge cloud and an internet cloud.

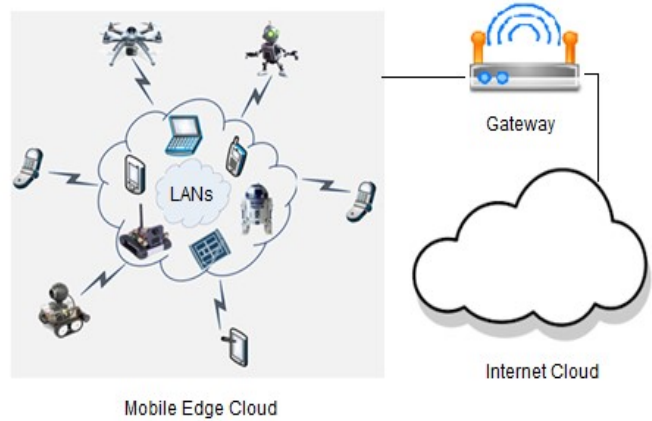

Figure 3: An illustration of mobile edge cloud

The architecture of mobile edge cloud (MEC) system is given in Figure 4. The architecture is organized into four layers: IoT Applications, MEC Middleware, Integrated Routing and Network Management, and Nodes layer. IoT Application layer includes applications such as smart home and mobile intelligent video surveillance. MEC Middleware layer is responsible for resource and task management, resource discovery, monitoring and allocation, task migration, failure management, and communication management. In addition, it hides all the complexities and provides a single system image to user and applications running on the system. Integrated Routing and Network Management layer provides a distributed network infrastructure for communication between heterogeneous devices across the network. Nodes layer includes heterogeneous mobile and stationary nodes. The key components of MEC Middleware and Integrated Routing and Network Management layers are described below.

*A. MEC Middleware layer*

Resource management system is responsible for the management of distributed computing and communication resources of mobile edge cloud. It includes several sub-components such as resource discovery and monitoring system, resource scheduling and task distribution system, and task migration system. Resource discovery and monitoring system is responsible for discovery of new resources and monitoring of existing resources. Resource scheduling and task distribution system concerns the allocation of application task to suitable computing nodes and distribution of tasks to selected nodes [12] [13]. Task migration system migrates task to suitable nodes in order to improve resource utilization, balance the load, reduce energy consumption or improve application performance [2].

Mobility management system maintain the history of user's mobility patterns which is used to predict the next location of mobile user or the future network topology [13].

This information is used by resource management system to reduce resource discovery and monitoring cost or to make effective and robust resource allocation decisions [12] [13]. The routing manager at Integrated Routing and Network Management layer also utilizes this information in order to select stable and robust communication links.

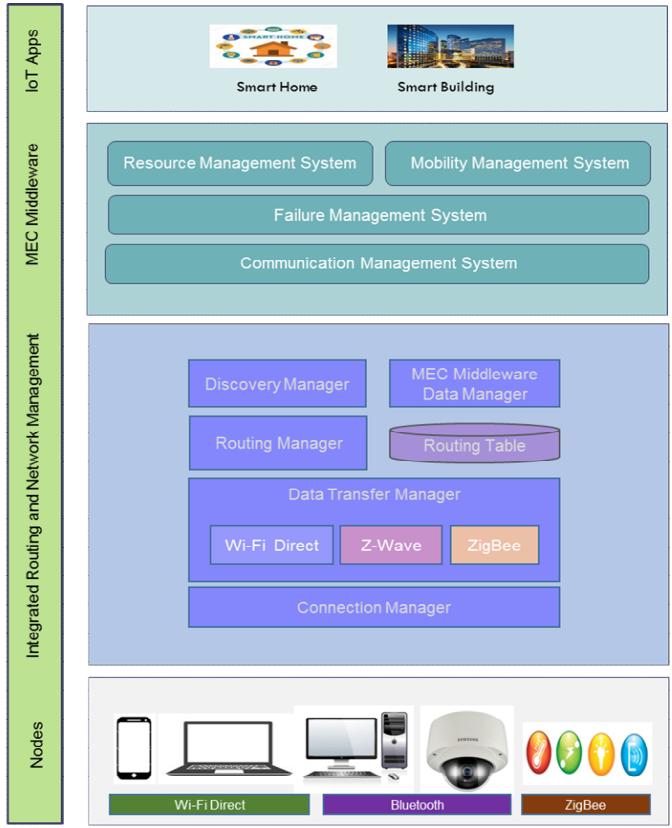

Figure 4: Mobile edge cloud architecture

Failure management system employs failure detection and avoidance mechanisms [13]. It relies on information supplied by mobility management system to avoid a failure due to node mobility and uses a resource monitoring system to detect a node or task failure. In response, it triggers task reallocation or migration process for reallocation or migration of an affected task. The decision of reallocation or migration of a task is taken based on the progress of application task.

Communication management system enables middleware components and tasks to communication with each other across the network. It hides all the complexities and provides a simple and uniform interface for the communication. Communication management system also serves as an interface between MEC Middleware layer and Integrated Routing and Network Management layer.

### B. Integrated Routing and Network Management layer

Discovery Manager performs following key tasks. (a) Discover devices in a communication range. (b) Periodically broadcast information stored in routing table to neighbor nodes. (c) Insert or update routing information received during the discovery process. Discovery Manager uses Data Transfer Manager to send, receive and broadcast discovery packets. It also communicates to Routing Table to manage routing information.

Routing Manager selects a route to a destination as per application requirement. For real-time applications it selects a route with minimum latency whereas for non-real-time applications it selects a route which consumes less energy. Routing Manager interacts with Routing Table to access routing information, Data Transfer Manager to send data to network or receive data from the network, and MEC Middleware Data Manager to send data to an application or receive data from an application for the transmission across the network.

MEC Middleware Data Manager hides all the network complexity and provides a simple and easy to use interface to MEC Middleware for transmission of data across the network.

Data Transfer Manager performs following key tasks. (a) Provide an interface to a respective wireless communication technology such as Wi-Fi Direct Z-Wave, and ZigBee. (b) Send and receive all type of packets such as control packets and data packets related to an application. It hides network heterogeneity and provides a uniform interface to all the components at Integrated Routing and Network Management layer.

Connection Manager is responsible for establishing or terminating a connection to a device. It is used by Data Transfer Manager to establish or terminate a connection.

RoutingTable stores routing and quality of service related information of discovered devices. The information is collected by network discovery and monitoring system.

## V. STATE-OF-THE-ART

The research on mobile edge cloud and similar systems is still in an initial phase. Several systems [5-7] [23-30] and schemes [8-12] [14-22] have been proposed to integrate mobile devices with internet cloud computing systems and address the issues such as resource allocation, load balancing, node mobility, energy management, and task failure. For example, authors of [33] have developed a system that allow a mobile user to execute computationally intensive tasks on an internet cloud. A similar system is developed in [31] [32] and [35]. Authors in [34] have proposed a three layer system architecture in which mobile nodes are connected to a cloudlet via Wi-Fi and cloudlet is connected to an internet cloud via a satellite.

The authors of [17] proposed a distributed resource allocation scheme which employs a proactive and reactive fault tolerance mechanism and supports redundant execution of tasks to address task failure. The scheme proposed in [18] utilizes a manager–worker model to distribute tasks and supports application-controlled migration to address failure due to low battery power. The problem of energy constrained scheduling is addressed in [19], where the authors investigated energy minimization and grid utility optimization problems. The scheme proposed in [14] also focuses on processing energy consumption while in [15] the cooperation among mobile devices is investigated to balance the energy consumption and the computation workload. To select the most suitable node for task execution, the authors of [16] proposed a scheme that utilizes a delayed reply mechanism in which a more resourceful node replies earlier than a less resourceful node. The scheme also provides load balancing and scalability. The scheme in [20] addresses node mobility by profiling regular movements of the user over time. Six online and batch scheduling heuristics are proposed in [21] to offload computationally intensive independent tasks on mobile ad hoc cloud. MinHop allocates task to a node accessible through minimum number of hops. Minimum Execution Time with Communication algorithm selects a node which takes least amount of time to execute a task whereas Minimum Completion Time with Communication algorithm assigns task to a node with minimum expected completion time. Both online scheduling algorithms take into account computation and communication cost but former focuses on execution time while later one on completion time. For batch scheduling, MinMinComm and MaxMinComm algorithms are proposed. The proposed scheduling heuristics are based on traditional MET, MCT, MinMin, and MaxMin heuristics and are extended to consider communication cost when estimating task completion time. The scheme proposed in [22] performs task allocations under varying assumptions about the connectivity environment. For example, in the ideal network environment where the future contact can be accurately predicted, the authors proposed a greedy task allocation algorithm that iteratively chooses the destination node for every task with the minimum task completion time.

In [8], various job stealing techniques such as random stealing, best rank aware stealing and worst rank aware stealing based on a centralized architecture were developed to reduce the processing energy consumption. The problem of unpredictable network connectivity, node mobility, energy consumption, and device failure has been addressed in [9]. To deal with uncertainty, an idea of application waypoints has been introduced [10] in which service provider executing application task reports to the broker with an estimate of residual task completion time. If the broker does not receive feedback about the estimated residual task completion time from the service provider at the specified waypoint, it marks the service provider as failed and assigns additional resources to take over the incomplete tasks. To reduce transmission energy consumption, we have proposed an energy efficient resource allocation scheme for allocation of tasks on mobile ad hoc grid [11]. The scheme is based on hybrid architecture that results in effective allocation and also reduces processing burden from a single node and communication cost associated with exchange of control information. To address node mobility problem, we proposed a two-phase resource allocation scheme in [12]. The scheme is divided into two phases. The first phase exploits the history of user mobility patterns to select nodes that provide long-term connectivity and the second phase takes into account the task and dependency types, and uses the distance information among the nodes selected in the first phase to reduce communication costs. An efficient and robust resource allocation scheme for allocation of interdependent tasks on a mobile ad hoc grid is proposed in [13]. The scheme aims to reduce the task completion time and the amount of energy consumed in the transmission of data.

VI. CONCLUSION

Numerous mobile cloud computing systems have been developed to support execution of data or computationally intensive applications or to provide vast amount of storage space and processing power. The existing systems are either developed for mobile ad hoc environments or pre-existing network infrastructure-based environments. In mobile edge cloud, multiple mobile and stationary devices interconnected through wireless ad hoc and pre-existing network infrastructure-based local area networks are combined to create a small cloud infrastructure. Mobile edge cloud system serve as a distributed computing and networking infrastructure for several internet of things and cyber physical system applications that require a vast amount of computing and storage resources. Mobile edge cloud also addresses the issues such as communication latency and transmission energy consumption associated with mobile infrastructure-based cloud. Compared to traditional mobile cloud computing systems, mobile edge cloud poses numerous complex challenges due to the heterogeneous computing environment, and heterogeneous and dynamic network environment. The real time requirements associated with internet of things and cyber physical system applications make the problem even more difficult.

In this paper, we have discussed the applications and challenges associated with design and development of mobile edge cloud system and have proposed an architecture which is divided into four layers: IoT Applications, MEC Middleware, Integrated Routing and Network Management, and Nodes layer. IoT Application layer includes applications whereas Nodes layer includes heterogeneous mobile and stationary nodes. MEC Middleware layer is responsible for resource and task management, resource discovery, monitoring and allocation, task migration, failure management, and communication management. Integrated Routing and Network Management layer provides a network infrastructure for communication between devices across the network. In future, we aim to develop algorithms and protocols for each component at MEC Middleware and Integrated Routing and Network Management layer.

ACKNOWLEDGMENT

This work was supported by Hankuk University of Foreign Studies Research Fund of 2017 and National Research Foundation of Korea (2017R1C1B5017629).